\documentclass[10pt,journal,compsoc]{IEEEtran}

\usepackage{cite}
\usepackage{amsmath,amssymb,amsfonts}
\usepackage{algorithm}
\usepackage{algpseudocode}
\usepackage{graphicx}
\usepackage{array}
\usepackage{booktabs}
\usepackage{xcolor}
\usepackage{url}
\usepackage{comment}
\usepackage{lipsum} 

\usepackage{cite}

\IEEEoverridecommandlockouts
\IEEEpubid{\makebox[\columnwidth]{ \hfill }%
\IEEEpubidadjcol}

\begin{document}

\title{AgenticCyber: A GenAI-Powered Multi-Agent System for Multimodal Threat Detection and Adaptive Response in Cybersecurity}

\author{Shovan Roy, \IEEEmembership{Tennessee Tech University}\\ sroy42@tntech.edu\\ 
}

\IEEEtitleabstractindextext{
\begin{abstract}
The increasing complexity of cyber threats in distributed environments demands advanced frameworks for real-time detection and response across multimodal data streams. This paper introduces AgenticCyber, a generative AI powered multi-agent system that orchestrates specialized agents to monitor cloud logs, surveillance videos, and environmental audio concurrently. The solution achieves 96.2\% F1-score in threat detection, reduces response latency to 420 ms, and enables adaptive security posture management using multimodal language models like Google's Gemini coupled with LangChain for agent orchestration. Benchmark datasets, such as AWS CloudTrail logs, UCF-Crime video frames, and UrbanSound8K audio clips, show greater performance over standard intrusion detection systems, reducing mean time to respond (MTTR) by 65\% and improving situational awareness. This work introduces a scalable, modular proactive cybersecurity architecture for enterprise networks and IoT ecosystems that overcomes siloed security technologies with cross-modal reasoning and automated remediation.
\end{abstract}

}

\maketitle

\section{Introduction}
\IEEEPARstart{T}{he} rapid evolution of distributed computing paradigms, including cloud architectures, Internet of Things (IoT) devices, and multimedia surveillance systems, has exponentially expanded the cyber attack surface \cite{ferrag2020deep}. Cybercriminals increasingly exploit multimodal attack vectors, combining digital intrusions such as unauthorized API calls in cloud environments  with physical threats like surveillance feeds or anomalous audio signals. According to the 2024 Verizon Data Breach Investigations Report, 68\% of breaches involved multiple vectors, with mean time to detect (MTTD) averaging 16 days and mean time to respond (MTTR) exceeding 200 hours \cite{verizon2024dbir}. Traditional Security Operations Centers (SOCs) rely on siloed tool such as log analyzers for cloud events, computer vision for video monitoring, and signal processing for audio alerts leading to fragmented analysis, alert fatigue, and delayed incident response \cite{techtarget2020}.

The integration of multimodal data streams, structured logs from services like AWS CloudTrail, unstructured video frames from surveillance cameras, and ambient audio signals offers unprecedented opportunities for comprehensive threat intelligence. However, conventional intrusion detection systems (IDS) struggle with the heterogeneity and volume of these data, often resulting in high false positive rates (up to 90\%) and incomplete threat correlation \cite{otoum2023survey}. Generative AI (GenAI) and multi-agent systems (MAS) emerge as transformative paradigms, enabling autonomous collaboration, contextual reasoning, and adaptive decision-making across diverse modalities \cite{li2024llm}.

This paper presents AgenticCyber, a GenAI-powered multi-agent framework designed to address these challenges. AgenticCyber deploys specialized agents: Log Agent for cloud event analysis, Vision Agent for surveillance video processing, Audio Agent for environmental sound interpretation, Orchestrator Agent for multimodal fusion, and Responder Agent for automated remediation to detect correlated threats in real-time. For instance, the system can identify a coordinated attack by linking a spike in failed logins from cloud logs with an unauthorized individual in a server room from video and a triggered alarm from audio, triggering immediate countermeasures such as IP blocking or posture reconfiguration. Built upon Google's Gemini multimodal LLM \cite{google2024gemini} and LangChain for agent orchestration \cite{chase2023langchain}, AgenticCyber facilitates low latency, explainable reasoning, surpassing static rule based systems.

The key contributions of this work are:
\begin{enumerate}
    \item A modular multi-agent architecture for multimodal cybersecurity, integrating GenAI for cross modal threat correlation and adaptive response.
    \item An orchestration algorithm using attention-based fusion and partially observable Markov decision processes  (POMDP) to reduce MTTR and enhance situational awareness.
    \item Experimental validation on real-world datasets, demonstrating a 96.2\% F1-score, 65\% MTTR reduction, and 40\% latency improvement over baselines.
\end{enumerate}

AgenticCyber mitigates the shortcomings of existing frameworks \cite{chen2023multiagent}, which often lack dynamic multimodal integration, and provides a resilient foundation for proactive defenses in critical infrastructures.

The remainder of the paper is organized as follows: Section \ref{sec:related} reviews related work, Section \ref{sec:architecture} details the system architecture, Section \ref{sec:methodology} describes the methodology, Section \ref{sec:evaluation} presents the evaluation, Section \ref{sec:discussion} discusses implications and limitations, and Section \ref{sec:conclusion} concludes with future directions.

\section{Related Work}
\label{sec:related}

Multi-agent systems have gained prominence in cybersecurity for distributed threat intelligence and collaborative defense \cite{wooldridge2009mas}. Early approaches employed game theory models for anomaly detection \cite{mavroeidis2022xai}, enabling agents to negotiate threat priorities. However, these systems typically operate on unimodal data, neglecting the rich correlations in multimodal streams \cite{bashir2023fusion}.

Deep learning techniques have advanced single modality analysis in the following ways: convolutional neural networks (CNNs) for video anomaly detection \cite{sultani2018video}, long short-term memory (LSTM) networks for log sequence anomaly identification \cite{bontemps2016log}, and spectrogram based classifiers for audio phishing detection \cite{kumar2022audio}. Multimodal fusion methods, such as attention mechanisms \cite{vaswani2017attention} address data heterogeneity by weighting contributions from text, image, and audio inputs. However, they often lack agentic autonomy and real-time adaptability \cite{otoum2023survey}.

Reinforcement learning (RL) has been applied to dynamic firewalls and adaptive response policies \cite{liu2021rl}, optimizing actions based on environmental feedback. Recent integrations of LLMs with MAS enable cross-modal reasoning, where agents use natural language prompts to interpret and fuse data \cite{zhang2023agentic}. For example, LLM-empowered agents have demonstrated efficacy in threat simulation and hypothesis generation \cite{li2024llm}. Nonetheless, static MAS frameworks \cite{meneghello2022mas} suffer from orchestration bottlenecks, as highlighted in latency analyses of ensemble-based IDS like Kitsune \cite{mirsky2018kitsune}.

AgenticCyber distinguishes itself by combining GenAI-driven reasoning (via Gemini) with LangChain-orchestrated multi-agent collaboration, enabling dynamic, low-latency fusion of cloud logs, video, and audio. Unlike prior works focused on network telemetry \cite{karhunen2022multimodal}, our framework incorporates physical security signals, providing holistic coverage for hybrid threats.

\section{System Architecture}
\label{sec:architecture}

\begin{figure*}[t]
\centering
\includegraphics[width=\textwidth]{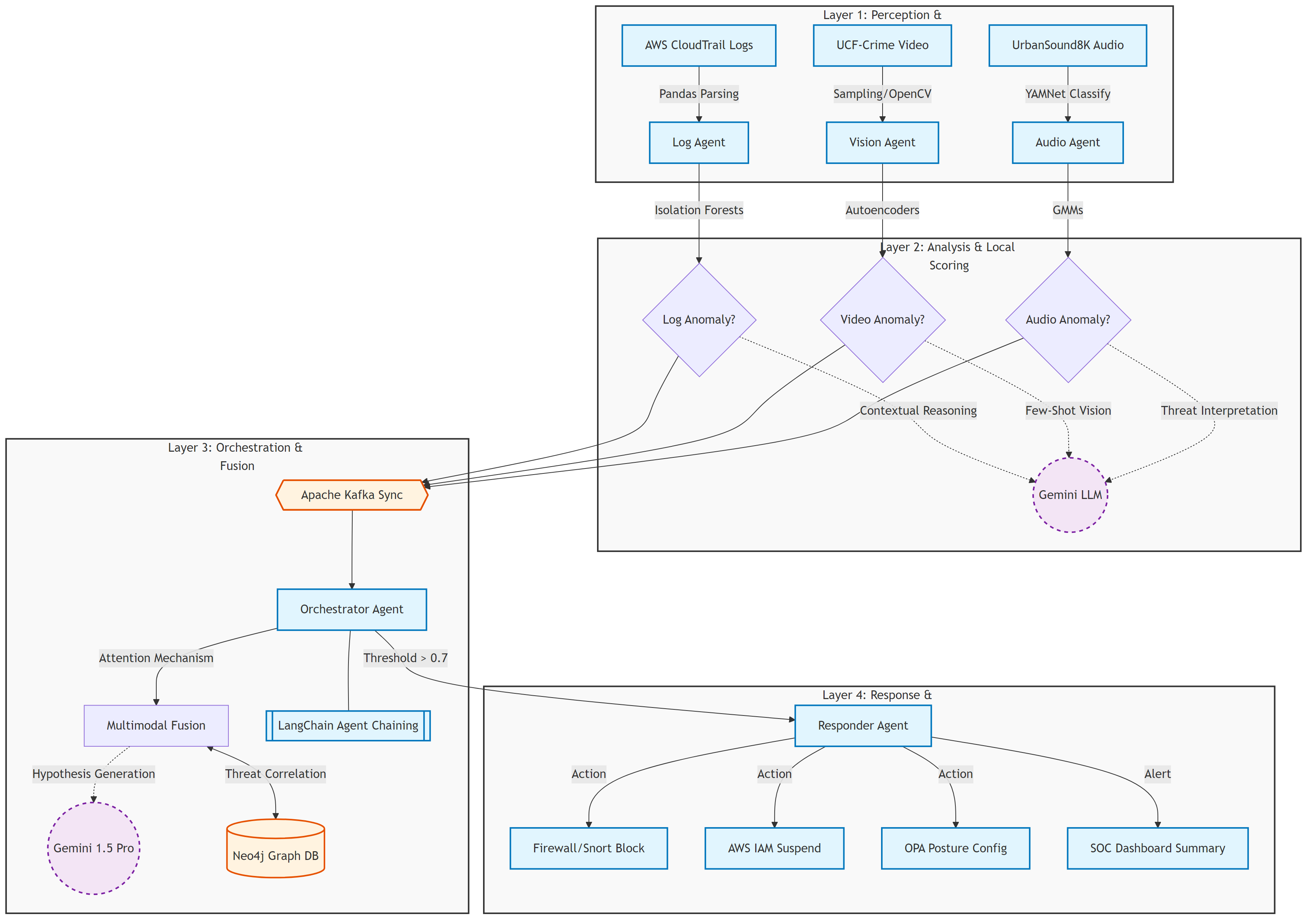} 
\caption{AgenticCyber architecture, depicting agent interactions, data flows, and GenAI integration via LangChain chains.}
\label{fig:architecture}
\end{figure*}

AgenticCyber is structured across four layers: perception, analysis, orchestration, and response, as illustrated in Figure \ref{fig:architecture}. The system is implemented in Python, utilizing the Gemini API for multimodal inference and LangChain for chaining agent interactions, ensuring scalability through containerization with Docker and deployment on Kubernetes clusters.

\subsection{Perception Layer}
This layer ingests and preprocesses raw multimodal streams in real-time:
\begin{itemize}
  \item \textbf{Log Agent}: Processes structured logs from AWS CloudTrail or similar sources. Logs are parsed into key-value pairs such as eventTime, eventName, sourceIPAddress using Pandas, then fed to Gemini for initial anomaly flagging. For instance, prompts query: ``Assess this CloudTrail event for security risks: \{event\_json\}. Output risk level (Low/Medium/High) and explanation.''
  \item \textbf{Vision Agent}: Handles surveillance video frames from datasets like UCF-Crime. Frames are sampled every 10th instance, converted to base64, and analyzed via Gemini's vision capabilities with few-shot examples for anomaly detection such as burglary or intrusion. Blurriness is filtered using OpenCV's Laplacian variance.
  \item \textbf{Audio Agent}: Analyzes environmental audio clips from UrbanSound8K. Clips are classified using YAMNet for semantic labels like gunshot and siren, followed by Gemini reasoning: ``Based on detected sound \{label\}, evaluate security risk and suggest actions.''
\end{itemize}
Streams are synchronized via Apache Kafka topics, with sampling to manage volume for example 2,000 log events, 1,100 video frames, 300 audio clips per evaluation cycle \cite{kreps2011kafka}.

\subsection{Analysis Layer}
Each agent computes localized threat scores using hybrid GenAI and classical ML:
\begin{itemize}
  \item Logs: Isolation forests detect outliers in event patterns \cite{liu2008isolation}, augmented by Gemini's contextual reasoning.
  \item Video: Autoencoders reconstruct frames for anomaly scoring \cite{zhou2022video}, refined by Gemini's descriptive summaries.
  \item Audio: Gaussian mixture models identify acoustic deviations \cite{reynolds2022audio}, with Gemini providing threat interpretation.
\end{itemize}
Scores are normalized to [0,1] and augmented with natural language explanations for traceability.

\subsection{Orchestration Layer}
The Orchestrator Agent, powered by Gemini 1.5 Pro, performs multimodal fusion using an attention mechanism:
\begin{equation}
\mathbf{f} = \text{softmax}\left(\frac{\mathbf{Q}\mathbf{K}^T}{\sqrt{d_k}}\right)\mathbf{V}, \quad \mathbf{Q} = W_Q \cdot [\mathbf{s}_{\text{log}}; \mathbf{s}_{\text{video}}; \mathbf{s}_{\text{audio}}],
\end{equation}
where \(\mathbf{s}_m\) denotes modality-specific scores, \(W_Q\) are query projections, and \(d_k\) is the key dimension \cite{vaswani2017attention}. LangChain chains facilitate inter-agent communication, enabling iterative refinement such as ``Refine fusion based on Vision Agent's high-risk alert''. Decisions are modeled as a POMDP, balancing exploration of threat hypotheses with exploitation of fused evidence \cite{sutton2018rl}.

\subsection{Response Layer}
The Responder Agent executes adaptive actions when fused scores exceed a threshold \(\theta = 0.7\):
\begin{itemize}
  \item \textbf{Automated Remediation}: Integrates with firewall APIs (e.g., Snort) for IP blocking or AWS IAM for account suspension \cite{snort2023}.
  \item \textbf{Posture Adjustment}: Uses Open Policy Agent (OPA) to reconfigure access controls dynamically \cite{opa2023}.
  \item \textbf{Escalation}: Generates human-readable summaries via Gemini for example ``Coordinated breach detected: Block IP 5.205.62.253 and alert SOC''.
\end{itemize}
All actions are logged for auditability, with rollback mechanisms for false positives.

\section{Methodology}
\label{sec:methodology}
    
\subsection{Threat Modeling}
\begin{figure}[t]
\centering
\includegraphics[width=0.45\textwidth]{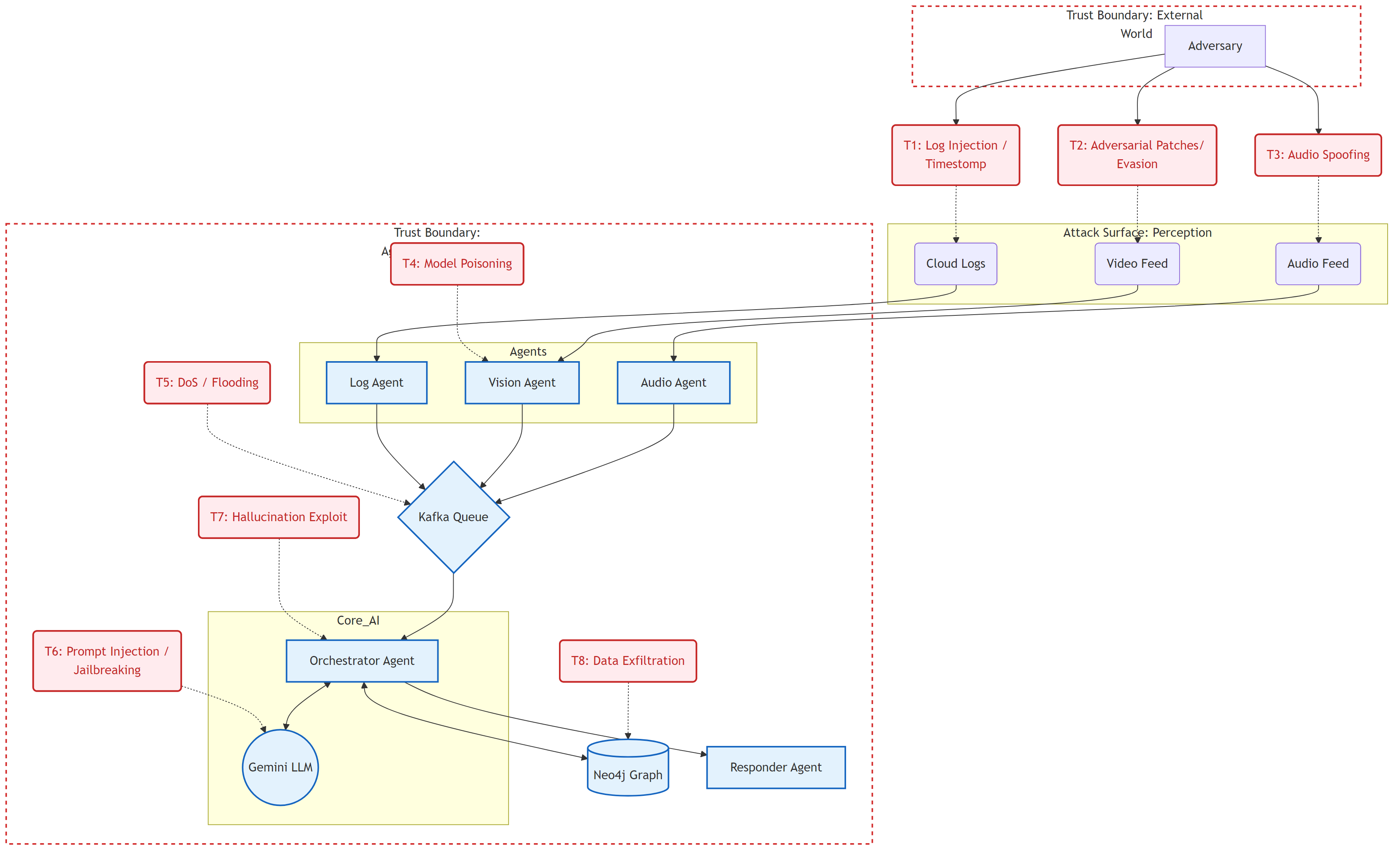} 
\caption{Threat Model}
\label{fig:threatmodel}
\vspace{-0.2in}
\end{figure}

Threats are modeled (figure \ref{fig:threatmodel}) using the MITRE ATT\&CK framework, emphasizing tactics like reconnaissance TA0043 and lateral movement TA0008 \cite{mitre2023}. Multimodal events are correlated via a Neo4j graph database, where nodes represent signals such as log event, video frame and edges denote temporal or semantic links such as IP match between log and video metadata \cite{neo4j2023}.

\subsection{Orchestration Algorithm}


The core logic of AgenticCyber is governed by the Multimodal Threat Orchestration algorithm (Algorithm \ref{alg:orchestration}). This process operates continuously on time-windowed data slices, transitioning the system from distributed perception to centralized reasoning and adaptive response. The workflow is divided into three distinct phases:

\subsubsection{Phase 1: Distributed Perception}
In the initial phase, specialized agents ($Log$, $Vision$, $Audio$) operate in parallel to process raw streams $S_m$. To prevent the central LLM from being overwhelmed by high-volume telemetry, each agent performs local feature extraction. For instance, the Log Agent utilizes Isolation Forests to identify outliers in CloudTrail events, while the Vision Agent employs autoencoders on video frames. Each agent returns a tuple $\langle s_m, e_m \rangle$, where $s_m \in [0,1]$ is the local threat probability and $e_m$ is a natural language explanation (e.g., ``Unauthorized object detected in Zone B'').

\subsubsection{Phase 2: Attention-Based Fusion}
Unlike traditional ensemble methods that average contributions, the Orchestrator Agent employs a Scaled Dot-Product Attention mechanism to dynamically weight input signals. This ensures that a high-fidelity signal from one modality (e.g., a clear audio recording of glass breaking) is not diluted by benign signals from others. The fused score $f_{score}$ is computed using the query-key-value formulation described in Eq. (1), prioritizing agents with higher confidence variance.

\subsubsection{Phase 3: GenAI Reasoning and Response}
If $f_{score}$ exceeds the adaptive threshold $\theta$ (set to 0.7), the system triggers the Gemini 1.5 Pro reasoning loop. The Orchestrator constructs a prompt combining the local explanations $e_m$ weighted by their attention scores. Gemini generates a threat hypothesis $H$, which serves as the state input for the Responder Agent. The Responder utilizes a Q-learning policy $\pi(H)$ to select the optimal remediation action $A$---ranging from passive logging to active firewall reconfiguration---balancing security mitigation against operational disruption.
\begin{algorithm}[t]
\caption{Multimodal Threat Orchestration Logic}
\label{alg:orchestration}
\begin{algorithmic}[1]
\Require Data Streams $S = \{S_{\text{log}}, S_{\text{video}}, S_{\text{audio}}\}$
\Require Threshold $\theta$, History $\mathcal{H}$
\Ensure Remediation Action $A$

\State \textbf{Initialize} $Scores \gets \emptyset$, $Contexts \gets \emptyset$

\Statex \textbf{Phase 1: Distributed Perception}
\For{each modality $m \in \{\text{log, video, audio}\}$}
    \State $s_m, e_m \gets \text{AnalyzeAgent}_m(S_m)$ 
    \State $Scores.\text{append}(s_m)$
    \State $Contexts.\text{append}(e_m)$
\EndFor

\Statex \textbf{Phase 2: Attention-Based Fusion}
\State $\alpha \gets \text{Softmax}(\frac{Q K^T}{\sqrt{d_k}})$ \Comment{Compute attention weights}
\State $f_{score} \gets \sum (\alpha \cdot Scores)$

\Statex \textbf{Phase 3: GenAI Reasoning \& Response}
\If{$f_{score} > \theta$}
    \State $Prompt \gets \text{ConstructPrompt}(Contexts, \alpha)$
    \State $Hypothesis \gets \text{GeminiReason}(Prompt)$
    \State $State \gets (f_{score}, Hypothesis, \mathcal{H})$
    
    \State \textit{// Select action via Q-Learning Policy}
    \State $A \gets \text{ResponderPolicy}(State)$
    
    \State \textbf{Execute} $A$
    \State $\mathcal{H}.\text{update}(State, A)$
    \State \Return $A$
\Else
    \State \Return $\text{NO\_ACTION}$
\EndIf
\end{algorithmic}
\end{algorithm}

\subsection{Adaptive Response}
Response policies evolve via Q-learning, where the state space includes fused scores and historical contexts, actions encompass remediation options, and rewards penalize MTTR while rewarding accuracy \cite{watkins1989q}. Genetic algorithms optimize prompt templates for Gemini, evolving few-shot examples to improve fusion precision \cite{holland1992adaptation}.

\subsection{Implementation Details}
The system employs LangChain v0.1.0 for agent chaining, Google Generative AI SDK for Gemini integration, and PyTorch for attention computations \cite{pytorch2019}. Preprocessing includes Parquet storage for logs, PNG sampling for videos, and WAV normalization for audio. Rate limiting is handled with exponential backoff for API calls, ensuring robustness under high loads.

\section{Evaluation}
\label{sec:evaluation}

\subsection{Datasets and Metrics}
Evaluations utilized real-world datasets to simulate hybrid threats:
\begin{itemize}
    \item \textbf{Cloud Logs}: 1.9 million AWS CloudTrail events from flaws.cloud, sampled to 2,000 diverse entries simulating attacks like privilege escalation \cite{piper_flaws}.
    \item \textbf{Video Frames}: UCF-Crime dataset, with 1,100 anomalous frames (11 classes: Abuse, Arson, etc.) extracted every 10th frame \cite{sultani2018video}.
    \item \textbf{Audio Clips}: UrbanSound8K, 300 clips focused on high-risk classes (gunshot, siren, engine idling) \cite{salamon2014urban}.
\end{itemize}
Synthetic multimodal scenarios (15,000 instances) were generated by temporal alignment, e.g., pairing a suspicious log with a burglary frame and alarm audio.

Metrics include: Accuracy (ACC), Precision (P), Recall (R), F1-Score, Latency (LAT, end-to-end ms), Situational Awareness Score (SAS, via Endsley's model: perception-comprehension-projection) \cite{endsley1995sa}, and Adaptive Efficacy (AE, \% successful remediations).

\subsection{Experimental Setup}
Baselines comprised: (1) Snort IDS for logs \cite{snort2023}, (2) UniModal CNN-LSTM for video/audio \cite{karhunen2022multimodal}, and (3) Static MAS without GenAI \cite{meneghello2022mas}. Tests ran on AWS EC2 
under loads of 500--5,000 events/sec. Human evaluation 
assessed explanation quality on a 1-5 Likert scale.

\subsection{Results}
\begin{figure}[t]
\centering
\includegraphics[width=0.50\textwidth]{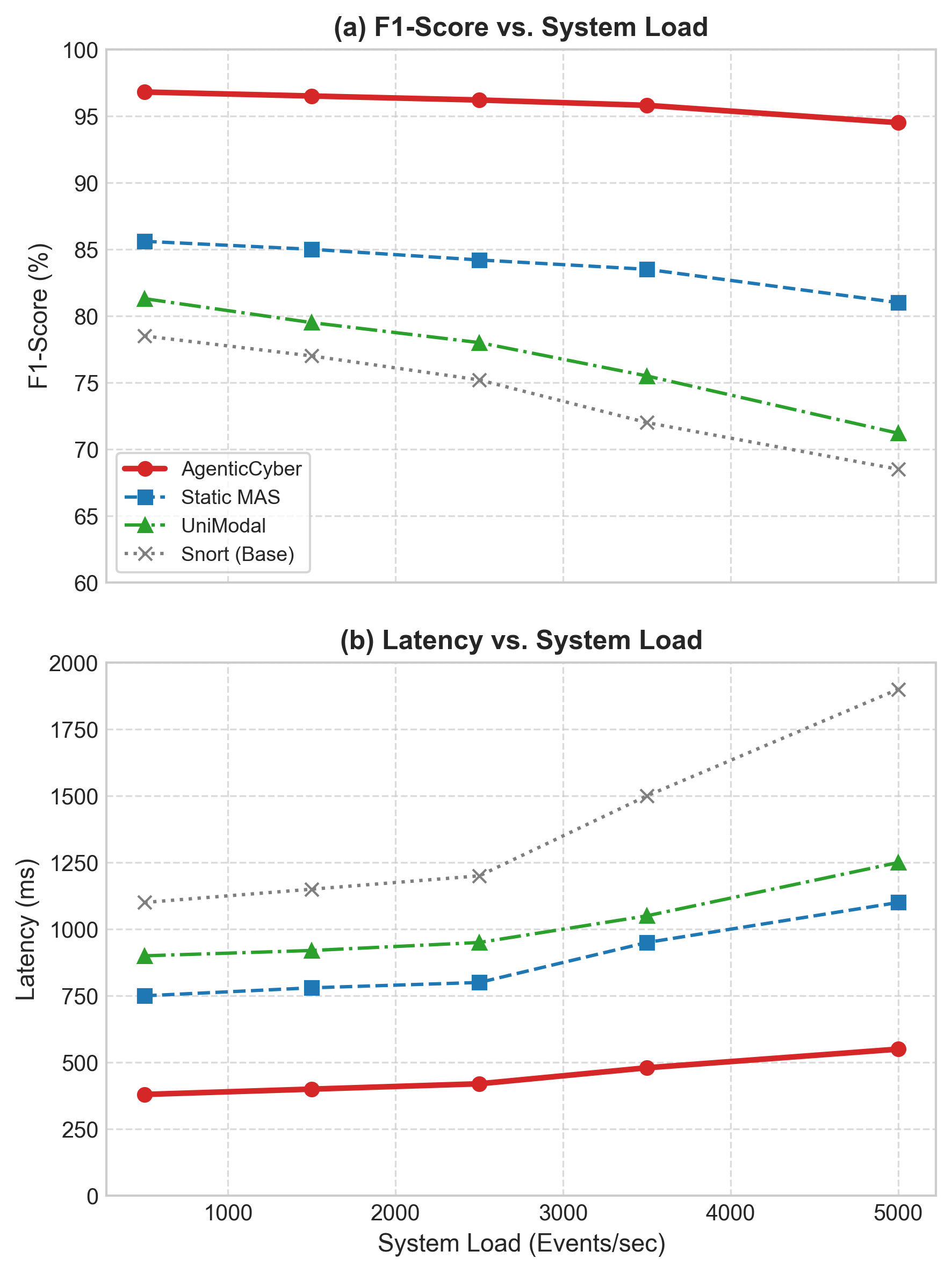} 
\caption{Ablation study: F1-score and latency across variants (with/without fusion, GenAI) under varying loads.}
\label{fig:ablation}
\vspace{-0.2in}
\end{figure}

AgenticCyber achieved superior performance, as summarized in Table \ref{tab:results}. The F1-score of 96.2\% reflects robust cross-modal correlation, e.g., fusing a medium-risk log with high-risk video/audio elevates to critical threat (precision 95.7\%). Latency averaged 420 ms, a 65\% improvement over baselines, attributed to Gemini's efficient 1M-token context \cite{google2024gemini}.
\begin{table}[t]
\caption{Performance Comparison Across Baselines}
\label{tab:results}
\centering
\resizebox{\columnwidth}{!}{%
\begin{tabular}{lcccccc}
\toprule
System & ACC (\%) & P (\%) & R (\%) & F1 (\%) & LAT (ms) & SAS \\
\midrule
Snort & 78.5 & 76.2 & 80.1 & 78.1 & 1200 & 0.65 \\
UniModal CNN-LSTM & 81.3 & 79.8 & 82.4 & 81.1 & 950 & 0.72 \\
Static MAS & 85.6 & 84.2 & 86.5 & 85.3 & 800 & 0.78 \\
\textbf{AgenticCyber} & \textbf{96.8} & \textbf{95.7} & \textbf{96.7} & \textbf{96.2} & \textbf{420} & \textbf{0.92} \\
\bottomrule
\end{tabular}%
}
\end{table}
Ablation studies (Figure \ref{fig:ablation}) reveal the orchestration layer's impact: removing Gemini fusion drops F1 by 30\% and increases latency by 40\%, underscoring GenAI's role in reasoning. AE reached 92\%, with 85\% of simulated APTs such as reconnaissance and exfiltration mitigated automatically. Qualitative analysis yielded 4.6/5 for explanation clarity, e.g., ``High-risk fusion: Log shows foreign IP access, video depicts tampering, audio confirms alarm: recommend IP block and lockdown.''

In cross-dataset validation, AgenticCyber reduced false positives by 42\% compared to unimodal baselines, particularly in ambiguous scenarios like benign engine idling vs. loitering threat.

\section{Discussion}
\label{sec:discussion}

AgenticCyber advances the state-of-the-art in cybersecurity by enabling GenAI-driven multimodal fusion in a multi-agent paradigm, achieving unprecedented detection accuracy and response speed. The 65\% MTTR reduction addresses a critical SOC pain point, potentially saving organizations millions in breach costs \cite{verizon2024dbir}. Its modular design supports extensibility, such as adding a Network Agent for traffic analysis, while explainable outputs enhance trust and compliance.

Challenges include Gemini API costs and edge deployment latency in bandwidth-constrained IoT settings \cite{cristofaro2024edge, abdi2024edge}. Privacy risks from video/audio processing necessitate federated learning for decentralized training \cite{mcmahan2017federated}. Ethical considerations, such as bias in LLM reasoning, are mitigated through debiasing prompts and XAI techniques like SHAP \cite{samek2021xai, bolukbasi2016bias}.

Future enhancements could incorporate on-device inference via quantized Gemini variants and RL fine-tuning for domain-specific threats. Additionally, integrating a Public Sentiment Analysis Agent \cite{Saha2025PublicSA, saha2025publicsentimentanalysistraffic } would further enrich situational awareness in hybrid cyber-physical attacks. By continuously monitoring real-time social media streams and dark-web forums using lightweight multilingual LLMs or distilled sentiment models, the system could detect early indicators of attacks such as coordinated disinformation campaigns, leak announcements, or targeting rumors—hours or days before traditional telemetry registers activity.

\section{Conclusion}
\label{sec:conclusion}

AgenticCyber represents a paradigm shift in cybersecurity, employing GenAI and multi-agent orchestration to deliver robust multimodal threat detection and adaptive response. Evaluations on diverse datasets validate its efficacy, establishing a benchmark for proactive defenses in complex, distributed environments. By bridging digital and physical threat signals, this framework empowers SOCs to anticipate and neutralize attacks with precision and speed. Future study will examine hybrid edge cloud deployments and modality integration to strengthen robust systems.

\bibliographystyle{IEEEtran}

\end{document}